\documentclass[twocolumn,twoside,slac_two]{revtex4}
\usepackage{graphicx}
\usepackage{fancyhdr}
\begin{document}

\title{Flaring $\gamma$-ray emission from high redshift blazars}

%

\author{M. Orienti, F. D'Ammando, M. Giroletti, D. Dallacasa, T. Venturi}
\affiliation{INAF-IRA, via Gobetti 101, 40129 Bologna, Italy}
\author{J. Finke}
\affiliation{US Naval Research Laboratory, Code 7653, 4555 Overlook Avenue SW, Washington, DC 20375-5352, USA }

\author{M. Ajello}
\affiliation{Department of Physics and Astronomy, Clemson University,
  Clemson, SC 29634, USA}  

\begin{abstract}

High redshift blazars are among the most powerful objects in the
Universe. Although they represent a significant fraction of the 
extragalactic hard X-ray sky, they are not commonly detected in
$\gamma$-rays. High redshift (z$>$2) objects represent $<$10\% of the
AGN population observed by {\it Fermi} so far, and $\gamma$-ray 
flaring activity from these sources is even more uncommon. 
The characterization of the radio-to-$\gamma$-ray properties of high 
redshift blazars represent a powerful tool for the study of both the
energetics of such extreme objects and the Extragalactic Background
Light. We present results of a multi-band campaign on TXS\,0536+145, 
which is the highest redshift flaring $\gamma$-ray blazar
detected so far. At the peak of the flare the source reached an 
apparent isotropic $\gamma$-ray luminosity of 6.6$\times$10$^{49}$
erg/s, which is comparable with the luminosity observed from the most 
powerful blazars. The physical properties derived from the 
multi-wavelength observations are then compared with those shown by
the high redshift population. In addition preliminary results from the
high redshift flaring blazar PKS\,2149-306 will be discussed. 

\end{abstract}

\maketitle

\thispagestyle{fancy}


\section{Introduction}

The population of high redshift ($z>2$) blazars represents a small fraction
($<$10\%) of the extragalactic $\gamma$-ray sky. They are mainly
associated with flat spectrum radio quasars (FSRQ), although a few BL
Lacs with $z>2$ are present in the third catalog of active galactic
nuclei detected by the Large Area Telescope (LAT) on board the {\it Fermi}
satellite after the first four years of 
scientific observations \cite{ackermann15}. The number counts drop
when higher energies are considered. In the first LAT catalog of
$\gamma$-ray sources above 10 GeV (1FHL) only seven objects with $z>2$
are detected \cite{ackermann13}.\\  
Although the detection of high redshift blazars during a $\gamma$-ray
flare is even more uncommon, the characterization of the
radio-to-$\gamma$-ray properties of high  
redshift blazars represent a powerful tool for the study of both the
energetics of such extreme objects and the Extragalactic Background
Light (EBL). During $\gamma$-ray flaring episodes the spectra of 
FSRQ sometimes show a moderate hardening \cite{pacciani14}, 
allowing us to explore energies that are usually strongly 
attenuated due to the intrinsic source spectrum. \\
So far, 10 blazars at $z > 2$ have been detected during $\gamma$-ray
flaring activity. Among these objects there are TXS\,0536+145 at
$z=2.69$, and PKS\,2149-306 at $z=2.34$. TXS\,0536+145 was not part of
the {\it Fermi}-LAT first (1FGL) and second source (2FGL) catalogs \cite{abdo10b,nolan12}, indicating its low
activity state during the first two years of {\it Fermi}-LAT
observations. On 2012 March 22 it underwent a $\gamma$-ray flare,
becoming the $\gamma$-ray flaring object at the highest redshift
observed so far \cite{orienti14}.\\
PKS\,2149-306 was detected by {\it Fermi}-LAT in a flaring state on
2013 January 4 \cite{dammando13}, with a daily $\gamma$-ray flux about 25 times
higher than the average source flux reported in the 2FGL
catalog \cite{nolan12}. \\
The high activity states observed in both sources triggered
multiwavelength monitoring observations aimed at characterizing the
variability in the various bands of the electromagnetic spectrum and
at determining the spectral energy distribution of these extreme
objects.\\

\section{TXS\,0536+145}

\subsection{{\it Fermi}-LAT data}

\begin{figure}
\begin{center}
\includegraphics{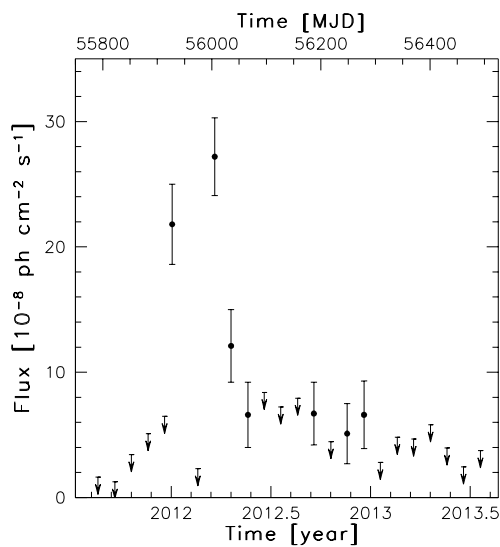}
\vspace{7cm}
\caption{Integrated {\it Fermi}-LAT light curve (0.1-100 GeV) of
  TXS\,0536+145 between 2011 August 4 and 2013 August 4 with 1-month
  time bins. Adapted from \cite{orienti14}.}
\label{fermi_0536}
\end{center}
\end{figure}

We analyzed {\it Fermi}-LAT data collected during the first five years
of scientific observations, from 2008 August 4 (MJD
54682) to 2013 August 4 (MJD 56508). 
We considered an energy range between 0.1 and 100 GeV,
and we followed the standard
LAT analysis procedures (for more details see \cite{orienti14}).\\
TXS\,0536+145 was not detected during the first two years of
observations. The 2$\sigma$ upper limit estimated over this period is
10$^{-8}$ ph cm$^{-2}$ s$^{-1}$. During the third and fourth years of
observations, the source was detected with a flux of
(4.2 $\pm$ 0.6)$\times$10$^{-8}$ ph cm$^{-2}$ s$^{-1}$ and a photon
index $\Gamma=2.37 \pm 0.09$. On 2012 March 22 the source was observed
during a $\gamma$-ray flare, when it reached a flux of (1.0 $\pm$ 0.3)$\times$10$^{-6}$ ph cm$^{-2}$ s$^{-1}$ and a photon
index $\Gamma=2.05 \pm 0.08$, indicating a hardening of the
spectrum. This flux corresponds to an apparent isotropic luminosity of
6.6$\times$10$^{49}$ erg/s.
Before this flare, the source was first detected in $\gamma$-rays
on 2012 January showing an enhancement of its high-energy activity,
but without reaching a similar peak flux (Fig. \ref{fermi_0536}). \\
The analysis of the {\it Fermi}-LAT data with E $>$ 10 GeV collected
between 2011 August and 2013 August could not detect the source at
such high energies. We evaluated the 2$\sigma$ upper limit as
9.3$\times$10$^{-11}$ ph cm$^{-2}$ s$^{-1}$ (assuming $\Gamma=2.37$).

\subsection{Radio properties}

\begin{figure}
\begin{center}
\includegraphics{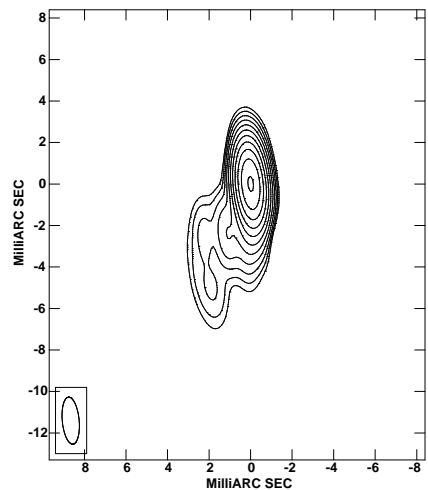}
\vspace{7cm}
\caption{VLBA image at 8.4 GHz of the source TXS\,0536+145. The peak
  flux density is 438.8 mJy/beam, while the first contour is 0.4
  mJy/beam and corresponds to three times the off-source noise level
  measured on the image plane. The contours increase by a factor of
  2. The restoring beam is plotted in the bottom-left corner. Adapted
  from \cite{orienti14}.}
\label{morpho}
\end{center}
\end{figure}

\begin{figure}
\begin{center}
\includegraphics{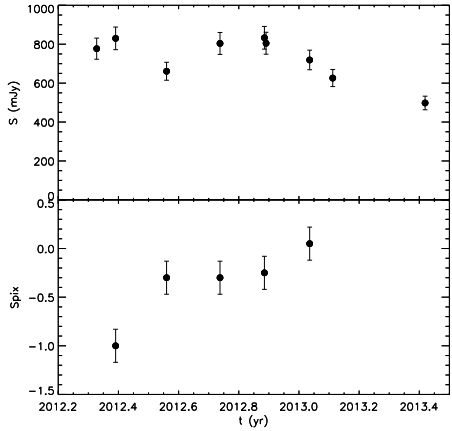}
\vspace{6cm}
\caption{Light curve at 15 GHz ({\it top}) and the spectral index computed
  between 8.4 and 15 GHz ({\it bottom}) for TXS\,0536+145. Adapted
  from \cite{orienti14}.}
\label{radio}
\end{center}
\end{figure}

Monitoring campaigns of TXS\,0536+145 
with the Very Long Baseline Array (VLBA) at 8.4,
15, and 24 GHz, and with the 
European VLBI Network (EVN) at 22 GHz were triggered by the
$\gamma$-ray flare with the aim 
of studying changes in the parsec-scale structure and the flux density
variability related to the central region of the source. The
observations were performed between 2012 April and 2013 October. The
source has a core-jet structure (Fig. \ref{morpho}). The radio emission is
dominated by the compact bright core component, which accounts for about 90
per cent of the total flux density at 8.4 GHz, and about 95 per cent
at 15 and 24 GHz. The jet emerges from the main component with a
position angle of about 180$^{\circ}$, then at $\sim$ 1.5 mas
(i.e. $\sim$ 12 pc) it 
slightly changes orientation to about 160$^{\circ}$ and extends to
$\sim$ 6 mas
(i.e. $\sim$ 48 pc).\\  
The flux density variability is
ascribed to the core region, while the jet is not variable. The radio
light curves show a flux density increase about 2-3 months after the
$\gamma$-ray flare, with longer delay occurring at lower
frequencies. The spectral index of the core
computed between 8.4 and 15 GHz shows a softening of the spectrum from
$\alpha \sim -1.0$ just after the flare, to $\alpha \sim 0.1$\footnote{The
  radio spectral index is defined as $S_{\nu} \propto \nu^{-
    \alpha}$} a few months later (upper panel of Fig. \ref{radio}). 
The light curve at 15 GHz shows a possible double hump similar to that
observed in $\gamma$-ray light curve (bottom panel of Fig. \ref{radio}). \\
No new superluminal component was observed after the flare. This may
be related to the high redshift of the target. In 
fact, only superluminal components with a speed higher than 35$c$
would have been picked up during the 16-month monitoring campaign.\\

\subsection{{\it Swift} data and SED}

Triggered by the flaring activity, {\it Swift} observed TXS\,0536+145
a few days after the 2012 March $\gamma$-ray flare, and the source was
found in a 
high state in X-rays. The X-ray flux decreases of a factor of two a couple
of weeks after the peak. An additional observation was
carried out a few months later, when the source was in a similar low
activity state. In the past, the source was not detected by the ROSAT
all-sky survey. Therefore, this is the first detection of
TXS\,0535+145 in X-rays.\\
Due to severe Galactic absorption and the short exposures, the source
was not detected by UVOT in any filter. \\
The hard X-ray flux of this source turned out to be below the
sensitivity of the BAT instrument for such short exposures, and
therefore the source was not detected. The source was not present in
the {\it Swift} BAT 70-month hard X-ray catalogue \cite{baumgartner13}.\\
The spectral energy distribution (SED) of TXS\,0536+145 in flaring activity
is well fitted by a synchrotron/external Compton model where the seed
photons upscattered to high energies may be those from the dusty torus
(see e.g., \cite{orienti14}). Due to the rather poor optical coverage,
the model parameter are not well constrained.\\

\section{PKS\,2149-306}

\begin{figure}
\begin{center}
\includegraphics{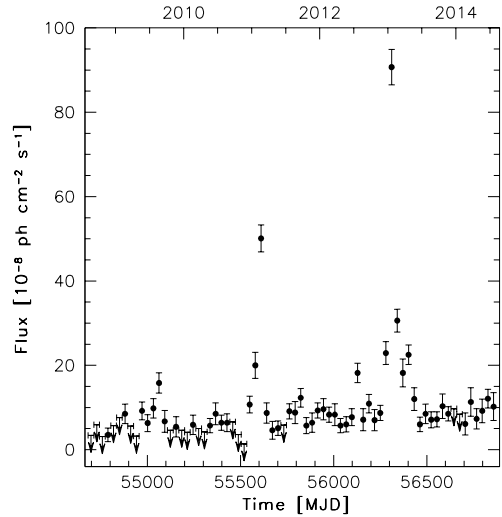}
\vspace{7cm}
\caption{Integrated {\it Fermi}-LAT light curve (0.1-100 GeV) of
  PKS\,2149-306 between 2011 August 4 and 2013 August 4 with 1-month
  time bins. Adapted from \cite{dammando15}.}
\label{fermi_2149}
\end{center}
\end{figure}

The FSRQ PKS\,2149-306 was observed by {\it Fermi}-LAT during a
flaring episode on 2013 January 4 and preliminary results were
reported in \cite{dammando13}. This source was part of the 2LAC
\cite{ackermann11}, indicating on average a higher level of
$\gamma$-ray activity 
with respect to TXS\,0536+145. \\
We analyzed {\it Fermi}-LAT data collected during the first six years
of scientific observations, from 2008
August 4 (MJD 
54682) to 2014 August 4 (MJD 56873). 
As in the case of TXS\,0536+145, we considered an energy range between
0.1 and 100 GeV, and we followed the standard LAT analysis procedures.\\
The source was
clearly detected by {\it Fermi}-LAT for most of the period with
one-month integration time (Fig. \ref{fermi_2149}). A first
significant increase of activity was observed in 2011 February. During
the strong flaring activity observed in 2013 January, the source
reached a daily peak flux of (3.0$\pm$0.4)$\times$10$^{-6}$ ph
cm$^{-2}$ s$^{-1}$, and showed a hardening of the spectrum. 
This value corresponds to an apparent peak
luminosity of 1.5$\times$10$^{50}$ erg/s. A dedicated multiwavelength analysis of
PKS\,2149-306 is ongoing aiming at studying the high-energy SED, also in
perspective of the next generation of high-energy telescopes.\\

\section{Discussion and conclusions}

High redshift flaring blazars are among the most luminous objects in
the Universe. The high redshift FSRQ TXS\,0536+145 and PKS\,2149-306
underwent a huge $\gamma$-ray flare, reaching an apparent isotropic
luminosity (0.1-100 GeV) of 6.6$\times$10$^{49}$ erg/s and
1.5$\times$10$^{50}$ erg/s, respectively. Such values are
comparable to the luminosity observed in the high-redshift 
gravitationally lensed blazar PKS\,1830-211 detected during a flare
($L_{\gamma} \sim 3\times 
10^{49}$ erg/s; \cite{abdo15}), as well as in the 
brightest flaring blazars, like
3C\,454.3 ($L_{\gamma} \sim 2\times 10^{50}$ erg/s; \cite{abdo11}),
PKS\,1510-089 ($L_{\gamma} \sim 4\times 10^{48}$ erg/s;
\cite{orienti13}), and PKS\,1622-297 ($L_{\gamma} \sim 4\times 10^{48}$
erg/s; \cite{mattox97}).\\ 
We compared the $\gamma$-ray properties of TXS\,0536+145 and
PKS\,2149-306 with those shown by the population of high redshift
($z>2$) $\gamma$-ray sources from the 2LAC \cite{ackermann11}. The
photon index and the luminosity in the low activity state of the
targets are in agreement with those of the other high-z
objects. During the flaring state both sources showed a hardening of
the spectrum. A similar behaviour was observed in the high-z flaring
blazar 4C+71.07 \cite{akyuz13}. \\
Despite the harder spectrum, no significant emission above 10 GeV is
observed for TXS\,0536+145. Although this value is consistent with current EBL
models (e.g. \cite{finke10}), the low statistics do not allow us
to attribute the spectral curvature to this effect \cite{orienti14}.
The improved sensitivity of the LAT at a few GeV with Pass 8 data will
be important for characterizing in more detail the $\gamma$-ray
spectrum of the high-redshift blazar population.

\begin{acknowledgments}
The \textit{Fermi} LAT Collaboration acknowledges generous ongoing support
from a number of agencies and institutes that have supported both the
development and the operation of the LAT as well as scientific data analysis.
These include the National Aeronautics and Space Administration and the
Department of Energy in the United States, the Commissariat \`a l'Energie Atomique
and the Centre National de la Recherche Scientifique / Institut National de Physique
Nucl\'eaire et de Physique des Particules in France, the Agenzia Spaziale Italiana
and the Istituto Nazionale di Fisica Nucleare in Italy, the Ministry of Education,
Culture, Sports, Science and Technology (MEXT), High Energy Accelerator Research
Organization (KEK) and Japan Aerospace Exploration Agency (JAXA) in Japan, and
the K.~A.~Wallenberg Foundation, the Swedish Research Council and the
Swedish National Space Board in Sweden.
Additional support for science analysis during the operations phase is
gratefully acknowledged from the Istituto Nazionale di Astrofisica in
Italy and the Centre National d'\'Etudes Spatiales in France. 
The VLBA is operated by the US National Radio Astronomy Observatory which is a facility of the National Science Foundation operated under a cooperative agreement by Associated Universities, Inc. The European VLBI Network is a joint facility of European, Chinese, South African, and other radio astronomy institutes funded by their national research councils.

\end{acknowledgments}


\end{document}